\documentclass[preprint,prb,
]{revtex4}
\usepackage{amsmath}
\usepackage{amssymb}
\usepackage{amsthm}
\newcommand{{\bff}}{\mbox{\boldmath$f$\unboldmath}}
\newcommand{{\bfF}}{\mbox{\boldmath$F$\unboldmath}}
\newcommand{{\bfA}}{\mbox{\boldmath$A$\unboldmath}}
\newcommand{\gradv}{\boldsymbol{\nabla}}
\def\v#1{{\bf#1}}

\begin{document}
\title{Can the Lorenz-gauge potentials be \\considered physical quantities?}
\author{Jos\'e A Heras}
\email{herasgomez@gmail.com}
\author{Guillermo Fern\'andez-Anaya}
\email{guillermo.fernandez@uia.mx}
\affiliation{Departamento de F\'isica y Matem\'aticas, Universidad Iberoamericana, Prolongaci\'on Paseo de la Reforma 880, M\'exico D. F. 01210, M\'exico}

\begin{abstract}
Two results support the idea that the scalar and vector potentials in the Lorenz gauge can be considered to be physical quantities: (i) they 
separately satisfy the properties of causality and propagation at the speed of light and not imply spurious terms and (ii) they can naturally be written in a manifestly covariant form. In this paper we introduce expressions for the Lorenz-gauge potentials at the present time in terms of electric and magnetic fields at the retarded time. These expressions provide a third result in favor of a physical interpretation of the Lorenz-gauge potentials: (iii) they can be regarded as causal effects of the observed electric and magnetic fields. 
\end{abstract}
\maketitle

\noindent{\bf 1. Introduction}
\vskip 10pt
The existence of a causal relation between sources and fields is universally accepted in
classical electrodynamics. 
This causal relation is
clearly expressed by the time-dependent generalizations of the Coulomb and Biot-Savart laws [1,2]. These generalized laws can be written in a form independent of specific units as follows [3,4]:
\begin{subequations}
\begin{align}
 \v E &\!=\frac{\alpha}{4\pi}\!\int \!d^3x'\bigg(\frac{\hat{\v R}}{R^2}[\rho]+\frac{\hat{\v R}}{Rc}\left[\frac{\partial \rho}{\partial t}\right]
-\frac{1}{Rc^2}\left[\frac{\partial \v J}{\partial t}\right]\bigg),\\
 \v B&\!=\frac{\beta}{4\pi}\!\int\! d^3x' \bigg([\v J]\times\frac{\hat{\v R}}{R^2 }+\bigg[\frac{\partial \v J}{\partial t}\bigg]\times\frac{\hat{\v R}}{R c}\bigg),
\end{align}
\end{subequations}
where $\hat{\v R}={\v R}/R =(\v x-\v x')/|\v x-\v x'|$; the square brackets $[\;]$ indicate that the enclosed quantity is to be evaluated at the retarded time $t'=t-R/c$ and the integrals are extended over all space.  For  SI units: $\alpha=1/\epsilon_0$ and $\beta=\mu_0$ and for Gaussian units 
$\alpha=4\pi$ and $\beta=4\pi/c.$ We should note that equation (1a) explicitly states that the present values of $\v E$ are determined by the retarded quantities $[\rho], [\partial \rho/\partial t]$ and $[\partial \v J/\partial t].$ But the property $[\partial/\partial t]= \partial[\;\:]/\partial t $ allows us to extract $\partial/\partial t$ outside the integrals and so the resulting expressions state that $\v E$ is ultimately caused  by $[\rho]$ and $[\v J].$ Similarly, equation (1b) states that $\v B$ is ultimately produced by $[\v J].$ In other words: the charge and current densities 
are implicit functions of the present time $t$ via the retarded time $t'=t-R/c$ and therefore, in the space integration of equations (1), we must evaluate the sources at the source point $\v x'$ and at the retarded time $t'=t-R/c$ and then the effects produced by these sources travel at the speed of light $c$ over the distance $R$ to arrive at the field point $\v x$ and at the present time $t.$ This retarded process agrees with the principle of causality, which briefly says that the causes (in this case $\rho$ and $\v J$) precede in time to their effects (in this case $\v E$ and $\v B$) [5,6]. %

In this paper we introduce a causal relation between the Lorenz-gauge
potentials $\Phi_L$ and $\v A_L$, which satisfy the Lorenz condition free of specific units: 
\begin{align}
\gradv\cdot\v A_L+\frac{\beta}{\alpha}\frac{\partial\Phi_L}{\partial t}=0,
\end{align}
and the electric and magnetic fields $\v E$ and $\v B$. We show that this causal relation is expressed in a form independent of specific units by the equations
\begin{subequations}
\begin{align}
 \Phi_L &\!=-\frac{1}{4\pi}\!\int \!d^3x'\bigg(\frac{\hat{\v R}}{R^2}\!\cdot\![\v E]\!+\!\frac{\hat{\v R}}{Rc}\!\cdot\!\left[\frac{\partial \v E}{\partial t}\right]\bigg),\\
 \v A_L&\!=\frac{1}{4\pi}\!\int \!d^3x' \bigg([\v B]\!\times\!\frac{\hat{\v R}}{R^2 }\!+\!\bigg[\frac{\partial \v B}{\partial t}\bigg]\!\times\!\frac{\hat{\v R}}{R c}\!-\!\frac{\beta}{\alpha}\frac{1}{R}\left[\frac{\partial \v E}{\partial t}\right]\bigg),
\end{align}
\end{subequations}
according to which the fields $\v E$ and $\v B$ produce the potentials  $\Phi_L$ and $\v A_L$
just in a similar way that equations (1) say that the densities $\rho$ and $\v J$ produce the fields $\v E$ and $\v B$. 
We also show that the causal link between fields and potentials given in equations (3) is physically consistent in the Lorenz gauge and not in other gauges. We emphasize that this result amounts to a physical interpretation of the Lorenz-gauge potentials. Finally, we briefly discuss other two arguments in favor of a physical interpretation these potentials. The first argument is based on the fact that the potentials $\Phi_L$ and $\v A_L$ separately satisfy the properties of causality and propagation at the speed of light and that such potentials do not imply spurious terms [7]. The second argument is based on the well-known result that the Lorenz-gauge potentials can naturally be written in a manifestly covariant form. 
\vskip 10pt

\noindent{\bf 2. The potentials $\Phi_L$ and $\v A_L$ as effects of the fields $\v E$ and $\v B $}
\vskip 10pt
The idea that the fields $\v E$ and $\v B $ produce the potentials  $\Phi_L$ and $\v A_L$ is strongly supported by equations (3). In fact, 
in equations (3) the fields $\v E$ and $\v B$ are implicit functions of the present time $t$ via the retarded time $t'=t-R/c$.  In the space integration of equations (3), we must evaluate the fields $\v E$ and $\v B$ at the point $\v x'$ and at the retarded time $t'=t-R/c$ and then the effects produced by these fields travel at the speed of light $c$ over the distance $R$ to arrive at the point $\v x$ and at the present time $t.$ This retarded process satisfies the principle of causality because the causes (in this case $\v E$ and $\v B$) precede in time to their effects (in this case $\Phi_L$ and $\v A_L).$ We also note that the result expressed in equations (3) is natural because of the following argument: if $\rho$ and $\v J$ produce $\v E$ and $\v B$ [equations (1)] and  $\v E$ and $\v B$ produce  $\Phi_L$ and $\v A_L$ [equations (3)] then the result that $\rho$ and $\v J$ produce  $\Phi_L$ and $\v A_L$ is naturally expected. In fact, this is a well-known result given by the equations
\begin{subequations}
\begin{align}
 \Phi_L &=\frac{\alpha}{4\pi}\int\!\! d^3x'\frac{1}{R}[\rho],\\
 \v A_L&=\frac{\beta}{4\pi}\int\!\!d^3x'\frac{1}{R} [\v J].
\end{align}
\end{subequations}

As emphasized by Jefimenko [5], not all relations connecting the fields $\v E$ and $\v B$
with the sources $\rho$ and $\v J$ can be considered causal equations. For example, both the Gauss law and the Ampere-Maxwell law:
\begin{subequations}
\begin{align}
\gradv\cdot\v E&=\alpha\rho,\\ 
\gradv\times \v B-\frac{\beta}{\alpha}\frac{\partial \v E}{\partial t}
&=\beta\v J,
\end{align}
\end{subequations}
cannot be considered to be causal relations because they connect quantities simultaneous in time. In a similar way,  the equations relating the Lorentz-gauge potentials with the electric and magnetic field:
\begin{subequations}
\begin{align}
-\gradv\Phi_L-\frac{\alpha}{\beta c^2}\frac{\partial \v A_L}{\partial t}
&=\v E,\\
\gradv\times\v A_L&=\v B,
\end{align}
\end{subequations}
cannot be considered as causal equations; they connect quantities simultaneous in time. 

It can be argued, however, that there exist equations simultaneous in time which can be defined to be causal relations in a different physical sense, which does not consider the causality principle. For instance, the mechanical relation $m\v a(t)=\v F(t)$ in which the force $\v F(t)$ acting on a particle of mass $m$ is defined to be the cause of the acceleration $\v a(t).$  The solution of this causal-by-definition relation is $m\v v(t)=\int_0^t\v F(t')dt'$ when $\v v(0)=0$. After performing the time integration, the velocity at the instant $t$ is in general an explicit function of its ``cause" at the instant $t$. 
On the other hand, equations (5) and (6) referee to a single time and also to a single space point but they are not causal relations by definition in the above sense. Integration of equations (5) and (6) lead in general to solutions at the instant $t$ expressed in terms of their sources at a different instant $t'$ [see equations (1) and (3)]. As pointed out by Jefimenko [4]: ``...an equation between quantities simultaneous in time and not separated in space cannot represent a causal relation between these quantities because, according to this principle [causality], the cause must precede its effect."

We now proceed to derive equations~(3). From equations~(4) and (5) we have
\begin{subequations}
\begin{align}
 \Phi_L &=\frac{1}{4\pi}\int\!\! d^3x'\frac{[\gradv'\cdot\v E]}{R},\\
 \v A_L&=\frac{1}{4\pi}\int\!\!d^3x'\bigg(\frac{[\gradv'\times\v B]}{R}- \frac{\beta}{\alpha}\frac{1}{R}\bigg[\frac{\partial \v E}{\partial t}\bigg]\bigg).
\end{align}
\end{subequations}
Making use of the identities:
\begin{subequations}
\begin{align}
\frac{[\gradv'\cdot\v E]}{R}\!=&\gradv\cdot\bigg(\frac{[\v E]}{R}\bigg)+
\gradv'\cdot\bigg(\frac{[\v E]}{R}\bigg)\nonumber \\=&
-\frac{\hat{\v R}}{R^2}\cdot[\v E]\!-\!\frac{\hat{\v R}}{Rc}\cdot\left[\frac{\partial \v E}{\partial t}\right]+
\gradv'\cdot\bigg(\frac{[\v E]}{R}\bigg)\\
\frac{[\gradv'\!\times\!\v B]}{R}\!=&\gradv\!\times\!\bigg(\frac{[\v B]}{R}\bigg)+
\gradv'\!\times\!\bigg(\frac{[\v B]}{R}\bigg)\nonumber\\
=&[\v B]\!\times\!\frac{\hat{\v R}}{R^2}\!+\!\left[\frac{\partial \v B}{\partial t}\right]\!\times \!\frac{\hat{\v R}}{Rc}+
\gradv'\!\times\!\bigg(\frac{[\v B]}{R}\bigg),
\end{align}
\end{subequations}
equations~(7) take the form
\begin{subequations}
\begin{align}
 \Phi_L &\!=-\frac{1}{4\pi}\!\int \!d^3x'\!\bigg(\frac{\hat{\v R}}{R^2}\!\cdot\![\v E]+\frac{\hat{\v R}}{Rc}\!\cdot\!\left[\frac{\partial \v E}{\partial t}\right]\bigg)\! +\!\frac{1}{4\pi}\!\int \!d^3x'\gradv'\!\cdot\!\bigg(\frac{[\v E]}{R}\bigg),\\
 \v A_L&\!=\frac{1}{4\pi}\!\int \!d^3x'\! \bigg([\v B]\!\times\!\frac{\hat{\v R}}{R^2 }\!+\!\bigg[\frac{\partial \v B}{\partial t}\bigg]\!\!\times\!\frac{\hat{\v R}}{R c}\!-\!\frac{\beta}{\alpha}\frac{1}{R}\!\left[\frac{\partial \v E}{\partial t}\right]\bigg)\!+\!\frac{1}{4\pi}\!\int \!d^3x'\gradv'\!\times\!\bigg(\frac{[\v B]}{R}\bigg).
\end{align}
\end{subequations}
The volume integrals in the last terms of equations (9) can be transformed into surface integrals,
which vanish at infinity under the assumption that both $\v E$ and $\v B$ fields are of order $O(|\v x|^{-2-\delta})$ where $\delta>0$ as $|\v x|$ tends to infinity. Such surface integrals also vanish if $\v E$ and $\v B$ are localized into a finite region of space.
Therefore, equations~(9) reduce to equations~(3). 

We note that equations (7) can alternatively be written as 
\begin{subequations}
\begin{align}
 \Phi_L &=\gradv\cdot\int\!\! d^3x'\frac{[\v E]}{4\pi R},\\
 \v A_L&=\gradv\times\int\!\!d^3x'\frac{[\v B]}{4\pi R}- \frac{\beta}{\alpha}\frac{\partial}{\partial t}\int\!\!d^3x'\frac{[\v E]}{4\pi R}.
\end{align}
\end{subequations}
These expressions can be obtained from equations (7) via an integration by parts using the property $[\partial/\partial t]= \partial[\;\:]/\partial t $ together with the identities (8) and assuming that both $\v E$ and $\v B$ are of order $O(|\v x|^{-2-\delta})$ where $\delta>0$ as $|\v x|$ tends to infinity. 

Equations (10) can also be written into a single manifestly covariant equation:
\begin{align}
A_L^\mu (x)=\partial_\mu\int d^4 x'G(x,x')F^{\mu\nu}(x'), 
\end{align}
where the Lorenz four-potential is defined by $A_L^\mu=\{(\beta c/\alpha)\Phi_L, \v A_L\}$; the electromagnetic field tensor $F^{\mu\nu}$ is defined by the polar components $F^{i0}=(\beta c/\alpha)(\v E)^i$ and the axial components $F^{ij}=-\varepsilon^{ijk}(\v B)_k$ where $(\v E)^i$ and $(\v B)_k$ represent the components of the electric and magnetic fields; $G(x,x')$ satisfies the wave equation $\partial_\nu \partial^\nu G=\delta(x-x')$ with $\delta(x-x')$ being the four-dimensional Dirac delta function and the integral is taken over all spacetime. The retarded form of $G(x,x')$ is given by $G(x,x')=\delta\{t'-t+R/c\}/(4\pi R)$. This form is not an explicit Lorentz invariant object. This objection can be avoided by replacing $G$ by the less-known Lorentz covariant retarded form $D(x,x')=[1/(2\pi)]\theta(x_0-x'_0)\delta[(x-x')^2],$ where $\theta$ is the theta function. 
Greek indices $\mu,\nu...$ run from 0 to 3; Latin indices $i,j...$ run from 1 to 3 the signature of the metric of the Minkowski spacetime is $(+,-,-,-);$  $x=(ct,\v x)$ is the field point and $x'=(ct',\v x')$ the source point; $\varepsilon^{ijk}$ is the totally antisymmetric three-dimensional tensor with $\varepsilon^{123} = 1.$ Summation convention on repeated indices is adopted;  derivatives in spacetime are defined by $\partial_\mu=\{(1/c)\partial/\partial t, \gradv\}$ and  $\partial^\mu=\{(1/c)\partial/\partial t,- \gradv\}.$ 

\vskip 10pt
\noindent{\bf 3. Other gauges}
\vskip 10pt
The causal relation between potentials and fields expressed in equations (3) is typical of the
Lorenz gauge. This causal relation is physically consistent in the Lorenz gauge and not in other gauges.
For example, we can show that there is no causal relation between the Coulomb-gauge potentials $\Phi_C$ and $\v A_C$ and the electric and magnetic fields $\v E$ and $\v B$. The Coulomb-gauge potentials satisfy the  
Coulomb condition 
\begin{align}
\gradv\cdot\v A_C=0,
\end{align}
and the following equations expressed in a form independent of specific units [7]:
\begin{subequations}
\begin{align}
\nabla^2\Phi_C&= -\alpha\rho,\\
\nabla^2{\v A_C}- \frac{1}{c^2}\frac{\partial^2\v A_C}{\partial t^2}&=-\beta\v J +
\frac{\beta}{\alpha}\frac{\partial}{\partial t}\gradv\Phi_C.
\end{align}
\end{subequations}
From the expression of the field $\v E$ given in terms of the Coulomb-gauge potentials it follows that $\gradv\Phi_C=-\v E -\{\alpha/(\beta c^2)\}\partial \v A_C/\partial t$, which is inserted into (13b) to obtain 
\begin{align}
\nabla^2{\v A_C}=-\beta\v J -\frac{\beta}{\alpha}\frac{\partial\v E}{\partial t}.
\end{align}
Making use of equations (5), (13a) and (14) we obtain the Poisson equations
\begin{subequations}
\begin{align}
\nabla^2\Phi_C&= -\gradv\cdot\v E,\\
\nabla^2{\v A_C}&=-\gradv\times\v B,
\end{align}
\end{subequations}
with the instantaneous solutions [8,9]
\begin{subequations}
\begin{align}
 \Phi_C(\v x,t) &=\frac{1}{4\pi}\int\!\! d^3x'\frac{\gradv'\cdot\v E(\v x,t)}{R},\\
 \v A_C(\v x,t)&=\frac{1}{4\pi}\int\!\!d^3x'\frac{\gradv'\times\v B(\v x,t)}{R}.
\end{align}
\end{subequations}
Using  the identities:
\begin{subequations}
\begin{align}
\frac{\gradv'\cdot\v E}{R}& \!=-\frac{\hat{\v R}}{R^2}\cdot\v E+
\gradv'\cdot\bigg(\frac{\v E}{R}\bigg), \\
\frac{\gradv'\times\v B}{R}&\!=\v B\times \frac{\hat{\v R}}{R^2}+
\gradv'\times\bigg(\frac{\v B}{R}\bigg),
\end{align}
\end{subequations}
equations~(16) take the form 
\begin{subequations}
\begin{align}
 \Phi_C &\!=-\frac{1}{4\pi}\int\!\! d^3x'\frac{\hat{\v R}}{R^2}\!\cdot\!\v E +\frac{1}{4\pi}\int\!\! d^3x'\gradv'\cdot\bigg(\frac{\v E}{R}\bigg),\\
 \v A_C&\!=\frac{1}{4\pi}\int\!\!d^3x'\:\v B\!\times\! \frac{\hat{\v R}}{R^2}\!+ \!\frac{1}{4\pi}\int\!\! d^3x'\:\gradv'\times\bigg(\frac{\v B}{R}\bigg).
\end{align}
\end{subequations}
The volume integrals in the last terms of equations~(18) can be transformed into surface integrals
which vanish at infinity under the assumption that both $\v E$ and $\v B$ are of order $O(|\v x|^{-2-\delta})$ where $\delta>0$ as $|\v x|$ tends to infinity. Therefore, equations~(18) reduce to 
\begin{subequations}
\begin{align}
 \Phi_C (\v x,t)&=-\frac{1}{4\pi}\int\!\! d^3x'\frac{\hat{\v R}}{R^2}\cdot\v E(\v x,t),\\
 \v A_C(\v x,t)&=\frac{1}{4\pi}\int\!\!d^3x'\:\v B(\v x,t)\times \frac{\hat{\v R}}{R^2}.
\end{align}
\end{subequations}
The Coulomb-gauge potential  $\Phi_C $ in equation (19a) cannot be considered to be an effect of the field $\v E$ because
both quantities are simultaneous in time which violates the principle of causality. By the
same reason, the potential  $\v A_C$ in equation (19b) cannot be regarded the effect of the field $\v B$.

Equations (3) as well as equations (19) are shown to be particular cases of a more
general relation existing between the potentials in the velocity gauge [7,10] and the electric
and magnetic fields. The velocity gauge is defined to be one in which the scalar potential
propagates with an arbitrary speed $v$. The velocity-gauge potentials $\Phi_v $ and $\v A_v$ expressed in a form independent of specific units satisfy the gauge condition
\begin{align}
\gradv\cdot\v A_v+\frac{\beta}{\alpha}\frac{c^2}{v^2}\frac{\partial\Phi_v}{\partial t}=0.
\end{align}
The velocity-gauge is actually a family of gauges which
contains the Lorenz gauge $(v = c)$ in equation~(2), the Coulomb gauge $(v = \infty)$ in equation~(12), and the recently introduced Kirchhoff gauge [13] 
$(v = ic)$ which satisfies $\gradv\cdot\v A_K-(\beta/\alpha)\partial\Phi_K/\partial t=0,$ where $\Phi_K$ and $\v A_K$ are the potentials in the Kirchhoff gauge. The velocity-gauge potentials satisfy the following set of coupled equations [7]:
\begin{subequations}
\begin{align}
\nabla^2\Phi_v- \frac{1}{v^2}\frac{\partial\Phi_v}{\partial t^2}&= -\alpha\rho,\\
\nabla^2{\v A_v}- \frac{1}{c^2}\frac{\partial\v A_v}{\partial t^2}&=-\beta\v J +
\frac{\beta}{\alpha}\bigg(1-\frac{c^2}{v^2}\bigg)\frac{\partial}{\partial t}\gradv\Phi_v.
\end{align}
\end{subequations}
These equations reduce to equations (13) when $v = \infty$. By inserting the quantity $\gradv\Phi_v=-\v E -\{\alpha/(\beta c^2)\}\partial \v A_v/\partial t$ into equation~(21b) we obtain 
\begin{align}
\nabla^2{\v A_v}- \frac{1}{v^2}\frac{\partial\v A_v}{\partial t^2}&=-\beta\v J -\frac{\beta}{\alpha}\frac{\partial\v E}{\partial t}+\frac{c^2}{v^2}\frac{\beta}{\alpha}\frac{\partial\v E}{\partial t} .
\end{align}
Making use of equations~(5), (21a) and (22) we obtain the wave equations
\begin{subequations}
\begin{align}
\nabla^2\Phi_v- \frac{1}{v^2}\frac{\partial\Phi_v}{\partial t^2}&= -\gradv\cdot\v E,\\
\nabla^2{\v A_v}- \frac{1}{v^2}\frac{\partial\v A_v}{\partial t^2}&=-\gradv\times\v B+ \frac{c^2}{v^2}\frac{\beta}{\alpha}\frac{\partial\v E}{\partial t},
\end{align}
\end{subequations}
with their retarded solutions
\begin{subequations}
\begin{align}
 \Phi_v &=\frac{1}{4\pi}\int\!\! d^3x'\frac{[\gradv'\cdot\v E]_v}{R},\\
 \v A_v&=\frac{1}{4\pi}\int\!\!d^3x'\bigg(\frac{[\gradv'\times\v B]_v}{R}- \frac{c^2}{v^2}\frac{\beta}{\alpha}\frac{1}{R}\bigg[\frac{\partial \v E}{\partial t}\bigg]_v\bigg),
\end{align}
\end{subequations}
where the square brackets $[\:\;]_v$ indicate that the enclosed quantity is to be evaluated at the retarded time $t'=t-R/v$.
Considering equations (24), using the identities (8) with the replacements $c\to v, [\;\;] \to [\;\;]_v$
and following the same procedure used to derive equations~(3), we obtain  
\begin{subequations}
\begin{align}
 \Phi_v &\!=-\frac{1}{4\pi}\int \!d^3x'\bigg(\frac{\hat{\v R}}{R^2}\cdot[\v E]_v+\frac{\hat{\v R}}{Rv}\cdot\left[\frac{\partial \v E}{\partial t}\right]_v\bigg),\\
 \v A_v&\!=\frac{1}{4\pi}\int \!d^3x' \bigg([\v B]_v\!\times\!\frac{\hat{\v R}}{R^2 }+\bigg[\frac{\partial \v B}{\partial t}\bigg]_v\!\!\times\!\frac{\hat{\v R}}{R v}\!-\!\frac{c^2}{v^2}\frac{\beta}{\alpha}\frac{1}{R}\left[\frac{\partial \v E}{\partial t}\right]_v\bigg).
\end{align}
\end{subequations}
These equations properly generalize equations (3) and (19). In fact, if $v = c$ then we have $[\v E]_v =[\v E], [\v B]_v = [\v B],  \Phi_v = \Phi_L, \v A_v = \v A_L$ and therefore equations (25) become equations (3). On the other hand, if $v =\infty$  then $[\v E]_v =\v E, [\v B]_v = \v B, \Phi_v = \Phi_C, \v A_v = \v A_C$ and therefore equations (25) become equations (19). 
According to equations (25), the fields $\v E$ and $\v B$ produce the potentials $\Phi_v$ and $\v A_v$.
This means that $\v E$ and $\v B$ in equations (25) are implicit functions of the present time $t$ via the retarded time $t'=t-R/v$, and therefore in the space integration of equations (25) we must evaluate the fields $\v E$ and $\v B$ at the point $\v x'$ and at the retarded time $t'=t-R/v$. The effects produced by these fields travel at the speed $v$ over the distance $R$ to arrive at the point $\v x$ and at the present time $t.$ This retarded process agrees with the principle of causality: the causes (in this case $\v E$ and $\v B$) precede in time to their effects (in this case $\Phi_v$ and $\v A_v).$ In particular, for $v > c$ (excluding $v = \infty$)  equations (25) define a causal relation between the potentials $\phi_v$ and $\v A_v$ and the fields $\v E$ and $\v B$. But special relativity demands that no physical propagation travel at speed greater than $c$. Therefore, equations (25) are physically acceptable only when $v=c$, that is, only in the case of the Lorenz potentials.

It is also of interest to consider the Poincare gauge proposed by Brittin $et$ $al$ [14], in which the scalar and vector potentials can be seen as a generalization of the well-known expressions 
$\phi=-\v x\cdot\v E_0$ and $\v A=\v B_0\times \v x/2$, where  $\v E_0$ and $\v B_0$ are static, uniform electric and magnetic fields (we are using SI units in this gauge).  In the general case, the Poincare-gauge potentials are given by [14]:  
\begin{subequations}
 \begin{align}
\phi(\v x,t)&= -\v x\cdot\int_0^1 d\lambda\;\v E(\lambda\:\v x,t),\\
\v A(\v x,t)&=\int_0^1\ \lambda\: d\lambda\:\v B(\lambda\:\v x,t)\times\v x,
\end{align}
\end{subequations}   
where  $\v E$ and $\v B$ are time-dependent electric and magnetic fields and the integrations involve only spatial variables.  As may be seen, in the Poincare gauge the connection between potentials and fields is naturally established. However,  the time $t$ is the same in both sides of Eqs.~(26). This means that the Poincare-gauge potentials  cannot be considered as causal effects of the electric and magnetic fields.

\vskip 10pt
\noindent{\bf 4. Discussion}
\vskip 10pt
The idea that the electromagnetic potentials have no physical meaning is strongly supported
by the fact that the scalar potential $\Phi$ in most gauges displays unphysical properties, like
instantaneous propagation in the potential $\Phi_C$ of the Coulomb gauge [7,10,11], imaginary propagation in the potential $\Phi_K$ of the Kirchhoff gauge [7,12],
arbitrary propagation (and in particular superluminal propagation $v>c$) in the potential $\Phi_v$ of the velocity gauge [7,10]  and no propagation in the potential $\Phi_s$ of the Coulomb static gauge [13]. 

Let us write $\Phi_\Omega$ where $\Omega=C,K,v,s$, that is, $\Phi_\Omega$ denotes some of the above mentioned scalar potentials. It follows that $-\gradv\Phi_\Omega$ inherits the unphysical properties of the scalar field $\Phi_\Omega$. The conceptual problem here is that the unphysical quantity  $-\gradv\Phi_\Omega$ is an explicit part of the physical field $\v E$, as may be seen in the following  expression of the electric field 
\begin{align}
\v E=-\gradv\Phi_\Omega-\frac{\alpha}{\beta c^2}\frac{\partial \v A_\Omega}{\partial t},
\end{align}
where $\v A_\Omega$ is the associated vector potential. For the gauges specified by the subscript $\Omega$ we can show the following result [7]:  
\begin{align}
-\frac{\alpha}{\beta c^2}\frac{\partial \v A_\Omega}{\partial t}= -\frac{\alpha}{4\pi}\int d^3x'\frac{[\gradv'\rho+(1/c^2)\partial\v J/\partial t]}{R}+ \gradv\Phi_\Omega.
\end{align}
Therefore from equations (27) and (28) we obtain the electric field with its experimentally verified properties of causality and propagation at speed of light:
\begin{align}
\v E=-\frac{\alpha}{4\pi}\int d^3x'\frac{[\gradv'\rho+(1/c^2)\partial\v J/\partial t]}{R}.
\end{align}
This means that the explicit presence of the unphysical quantity $-\gradv\Phi_\Omega$ inside the physical field $\v E$ given in equation (27) turns out to be irrelevant because such a quantity always is canceled by the component $+\gradv\Phi_\Omega$ contained in the term $\{\alpha/(\beta c^2)\}\partial \v A_\Omega/\partial t$ in equation (28). Therefore the term $-\gradv\Phi_\Omega$ in equation (27) constitutes a formal result of the theory with no physical meaning [7], that is, $-\gradv\Phi_\Omega$ is factually a spurious field. 

However, for the case of the Lorenz-gauge potentials we have  
\begin{subequations}
\begin{align}
-\gradv\Phi_L&= -\frac{\alpha}{4\pi}\int d^3x'\frac{[\gradv'\rho]}{R},\\
-\frac{\alpha}{\beta c^2}\frac{\partial \v A_L}{\partial t}&=-\frac{\alpha}{4\pi}\int d^3x'\frac{[(1/c^2)\partial\v J/\partial t]}{R}.
\end{align}
\end{subequations}
As may be seen, the quantity $-\{\alpha/(\beta c^2)\}\partial \v A_L/\partial t$ in equation (30b) does not contain spurious terms as in the case of the quantity $-\{\alpha/(\beta c^2)\}\partial \v A_\Omega/\partial t$
in equation (28). Equations in (30) display the property of causality and the property of propagation at the speed of light $c$. The latter property is consistent with 
the fact that both quantities $-\gradv\Phi_L$ and $-\{\alpha/(\beta c^2)\}\partial \v A_L/\partial t$ separately satisfy the wave equations 
\begin{subequations}
\begin{align}
\nabla^2(-\gradv\Phi_L)- \frac{1}{c^2}\frac{\partial^2}{\partial t^2}(-\gradv\Phi_L)&=\alpha\gradv\rho,\\
\nabla^2\bigg(\!\!\! -\frac{\alpha}{\beta c^2}\frac{\partial \v A_L}{\partial t}  \!\!    \bigg)- \frac{1}{c^2}\frac{\partial^2}{\partial t^2}
\bigg(\!\!\! -\frac{\alpha}{\beta c^2}\frac{\partial \v A_L}{\partial t}\!\!\bigg) &=\frac{\alpha}{c^2}\frac{\partial \v J}{\partial t}.
\end{align}
\end{subequations}
All of these results are characteristics of the Lorenz gauge. 
The physical character of the terms in equations (30) is strongly supported by the fact that their sum $-\gradv\Phi_L-\{\alpha/(\beta c^2)\}\partial \v A_L/\partial t$, namely, the electric field, is physically detectable. That $\Phi_L$ and $\v A_L$ satisfy the physical properties of causality and propagation at the speed of light and that these potentials do not imply spurious terms constitute two results that amount to a physical interpretation of the Lorenz-gauge potentials. 

A second argument supporting a physical interpretation of $\Phi_L$ and $\v A_L$ is based on the fact that these potentials can be written in a covariant form. The scalar and vector potentials form a four-vector which can be written as $A^\mu=\{(\beta c/\alpha)\Phi, \v A\}.$ The inhomogeneous Maxwell equations given in terms of $A^\mu$ can be written as $\partial_\nu\partial^\nu A^{\mu}-\partial^\mu\partial_\nu A^{\nu}=\beta J^\mu,$ where $J^\mu=\{c\rho, \v J \}$ is the four-current.  If we impose the Lorenz condition $\partial_\mu A_L^\mu=0,$ then the wave equation for the four-potential in this gauge reads 
$\partial_\nu\partial^\nu A_L^{\mu}=\beta J^\mu$ with its retarded solution given by 
\begin{align}
A_L^\mu (x)=\beta\int d^4 x'G(x,x')J^\mu(x').
\end{align}
The derivative of $A_L^\mu$ can be written as:
\begin{align}
\partial^\nu A_L^\mu (x)=\beta\int d^4 x'G(x,x')\partial'^\nu J^\mu(x'),
\end{align}
is also a Lorentz tensor. As may be seen, equation (33) does not contain spurious terms. The time and space components of equation (33) essentially  reproduce equations (30a) and (30b). The fact that $A_L^\mu$ is a Lorentz four-vector and the result that $A_L^\mu$ does not imply spurious terms are two covariant arguments for a physical interpretation of this four-potential.

\vskip 10pt
\noindent{\bf 5. Conclusion}
\vskip 10pt

In this paper we have introduced equations (3) which  causally connect the scalar and vector potentials in the Lorenz gauge with the electric and magnetic fields. We have noted that the Lorenz-gauge potentials are the only potentials propagating at the speed of light that can be considered as causal effects of the observed electric and magnetic fields. This result provide an argument in favor of a physical interpretation of  the Lorenz-gauge potentials. We have also emphasized other two arguments: (i) these potentials separately satisfy the properties of causality and propagation at the speed of light and not imply spurious terms and (ii) these potentials can naturally be written in a manifestly covariant form. Summarizing, in purely  classical electrodynamic considerations we have emphasized arguments in favor of a physical interpretation of the Lorenz-gauge potentials.

{}

\end{document}